\documentclass[12pt]{amsproc}

\theoremstyle{definition}

\theoremstyle{remark}

\numberwithin{equation}{section}

\usepackage{amsmath,amssymb,amsfonts}
\usepackage{graphicx}

\begin{document}

\title{On the spectral gap in Andreev graphs}

\author{Holger Flechsig}
\address{Fachbereich Physik, Freie Universit\"at Berlin, Germany}
\curraddr{Department of Physical Chemistry,
Fritz-Haber-Institute of the Max Planck Society,
Berlin, Germany}
\email{flechsig@fhi-berlin.mpg.de}
\author{Sven Gnutzmann}
\address{School of Mathematical Sciences, University of Nottingham, United Kingdom}
\email{sven.gnutzmann@nottingham.ac.uk}

%    General info
%\subjclass{Primary 54C40, 14E20; Secondary 46E25, 20C20}
%\date{January 1, 1994 and, in revised form, June 22, 1994.}


%\keywords{Differential geometry, algebraic geometry}

\begin{abstract}
  We introduce Andreev scattering 
  (electron-hole conversion at an interface 
  of a normal conductor to a superconductor)
  at the outer vertices of a quantum star graph 
  and examine its 
  effect on the spectrum. 
  More specifically we show that the density 
  of states in Andreev graphs
  is suppressed near the Fermi energy
  where a spectral gap may occur. 
  The size and existence of such a gap
  depends on one side on 
  the Andreev scattering amplitudes 
  and, on the other side,on the
  properties of the electron-electron 
  scattering matrix at the central vertex.
  We also show that the bond length 
  fluctuations have a minor effect
  on the spectrum near the Fermi energy.
\end{abstract}

\maketitle
% \begin{document}

% \title{The gap statististics in Andreev graphs}

% \author{Holger Flechsig\\
%   Institut f\"ur Theoretische Physik,\\
%   Freie Universit\"at
%   Berlin,\\ Arnimallee 14,\\ 14195 Berlin, Germany\\
%   flechsig@physik.fu-berlin.de\\[0.3cm]
%   and\\[0.3cm]
%   Sven Gnutzmann\\
%  School of Mathematical Sciences,\\
%   University of Nottingham,\\
%   University Park,\\ Nottingham , United Kingdom\\
% sven.gnutzmann@nottingham.ac.uk}

% \date{\today}

%\maketitle

%\begin{abstract}
%  We investigate hybrid superconducting-normalconducting
%  systems in the context of quantum graphs. 
%\end{abstract}

\section{Introduction and physical background}

The spectra of 
superconducting-normalconducting 
(SN) hybrid devices such as Andreev billiards 
has been a topic
in physics for some time (see the recent review 
\cite{beenakker} for an overview and further references).
An Andreev billiard consists of a finite 
normalconducting region with a boundary that is 
partly an interface to a superconductor.
An electron with energy near the 
Fermi energy cannot propagate into the 
superconductor and it is trapped
in a finite region similarly to a usual 
quantum billiard.
However, the interface to the superconductor 
leads to an effective coupling of electron and 
hole dynamics through Andreev scattering.
If an electron hits the interface to the 
superconductor it is reflected back into the 
billiard as a hole in the opposite direction of
the incoming electron (this has been called ``retroflection'' to distinguish it from the 
specular reflection at an interface to an 
insulator).
Physically the electron enters the superconductor
where it 
excites an electron-hole pair.
It then builds a Cooper 
pair with the electron while the hole is
ejected back into the billiard.
Quasiparticle excitations in such a device are 
described
by the Bogoliubov-de Gennes equation
\begin{equation}
  \begin{pmatrix}
    -\nabla^2 + V(\mathbf{x})-\mu & \Delta(\mathbf{x})\\[0.3cm]
    \Delta(\mathbf{x}) & \nabla^2-V(\mathbf{x})+\mu
  \end{pmatrix}
  \boldsymbol\psi(\mathbf{x})= E \boldsymbol\psi(\mathbf{x})\ .
  \label{eq:BdG}
\end{equation}
Here, $V(\mathbf{x})$ is a potential, 
$\mu\equiv k_F^2$ is the Fermi energy ($k_F$ is the Fermi wave number), and $E$ the energy 
measured from the Fermi level. 
The wave function 
$
  \boldsymbol\psi(\mathbf{x})=
  \begin{pmatrix} u(\mathbf{x}) \\ v(\mathbf{x}) 
  \end{pmatrix}
$
has two components: $u(\mathbf{x})$ is the electron 
component and $v(\mathbf{x})$ the hole 
component. The electron and hole 
components are coupled by the
pair potential $\Delta(\mathbf{x})$. 
The pair potential vanishes in a 
normalconducting region. 
In superconducting regions
$\Delta(\mathbf{x}) \neq 0$ and waves can only 
propagate for $E>|\Delta(\mathbf{x})|$.
In \eqref{eq:BdG} we have assumed 
time-reversal invariance which means
that there is no magnetic field and the pair potential is a 
real function. 
If $\boldsymbol\psi(\mathbf{x})=\begin{pmatrix} u(\mathbf{x}) \\ v(\mathbf{x}) \end{pmatrix}$
solves the Bogoliubov-de Gennes equation 
at energy $E$
then $\mathcal{C} \boldsymbol\psi(\mathbf{x})=\begin{pmatrix} -v(\mathbf{x})^* \\ u(\mathbf{x})^* \end{pmatrix}$
is a solution to energy $-E$. This shows 
that the spectrum
is symmetric around $E=0$ 
(that is around the Fermi energy) and 
reveals that
these systems belong to one of the seven
non-standard symmetry classes \cite{AZ} 
(the three standard or Wigner-Dyson 
symmetry classes classify quantum systems 
according to their behaviour under 
time-reversal and spin rotation). 
The negative spectrum describes
hole-like quasiparticle 
excitations (while the spectrum is negative, the physical
energy of such an excitation is positive, of course) and the positive
spectrum describes electron-like
quasiparticle excitations.\\
It has been known for some time that
Andreev reflections reduce the density 
of states near the Fermi energy and various
universality classes have been defined 
\cite{AZ,melsen,frahm,lodder,wiersig,cserti,
  cserti2,ossipov,selfdual} 
and related to quantum chaos.
Spectral gaps have been found in 
irregularly shaped Andreev billiards for which 
the classical dynamics of the normal billiard 
(with specular reflection at the 
superconducting interface) is chaotic. 
Though Andreev 
scattering destroys chaos the Lyapunov exponent 
can be felt if the mean time $t_A$ between two
subsequent 
Andreev reflections is large -- the gap
that was found is of the order $\hbar/t_A$
(in general much larger than the 
mean level spacing) \cite{melsen,frahm,lodder}. 
In contrast an integrable
normal billiard only leads to a suppressed density
of states near the Fermi energy \cite{cserti,cserti2}.
Other universality classes such as magnetic 
billiards may have a chaotic dynamics even in the presence of
Andreev reflections -- 
in that case the effect of
Andreev scattering is 
universally described by appropriate Gaussian random-matrix ensembles
\cite{AZ,selfdual}.
However the effect of Andreev scattering 
on the spectrum is less strong
and decays after a few mean level spacings away from
Fermi energy (which can only be seen 
after sufficiently averaging over system 
parameters). 
\\
Andreev scattering has been 
introduced to quantum maps \cite{jaquod} and
quantum graphs \cite{Andreevgraphs}.
The latter work investigated a class
of Andreev graphs that corresponds to 
chaotic combined electron-hole dynamics
with universal spectral statistics
for non-standard symmetry classes.
Chaos is however suppressed in
a non-magnetic Andreev 
billiard.
Here we will discuss 
the effect of Andreev reflections in
a different class of graphs that is more closely 
related to non-magnetic Andreev billiards.

In Section \ref{sec:lead}
we will start our discussion
with a simple model: a single lead with
one interface to a superconductor at one end.
This will serve as a motivation for our
construction of Andreev graphs in Section
\ref{sec:construction}. In the rest of Section
\ref{sec:gaps} we will discuss the spectra
of various ensembles of Andreev graphs -- 
partly analytically,
and numerically where analytical results
could not be obtained straight forwardly.

\section{A hybrid SN lead}
\label{sec:lead}

Most of the concepts we will use later 
are most easily 
introduced for a one dimensional electronic 
hybrid SN lead. 

\subsection{Scattering from a SN interface}

Let us start with discussing scattering from a 
SN interface
at $x=0$. Let us assume that the pair 
potential is constant on the 
superconducting half line $x>0$
where $\Delta(x)\equiv \Delta_0$ while 
$\Delta(x)\equiv 0$ in the normalconducting
half line  $x<0$.
At the interface we include a potential barrier 
$V(x)=V_0 \delta(x)$.
The scattering problem can
be solved exactly for any set of parameters.
%$\mu$, $\Delta_0$, $V_0$, and $E$. 
We are interested in the case
$E<\Delta_0$ in which there is no propagation 
in the superconducting half line.
On the normalconducting half line a 
scattering solution of the Bogoliubov-de Gennes equation 
takes the form
\begin{equation}
  \boldsymbol{\psi}_{\mathrm{NC}}(x)=
  \begin{pmatrix}
    a_{e, \mathrm{in}}\frac{\displaystyle e^{ik_e x}}{\displaystyle \sqrt{k_e}}\\[0.4cm]
    a_{h, \mathrm{in}}\frac{\displaystyle e^{-ik_hx}}{\displaystyle \sqrt{k_h}}
  \end{pmatrix}
  +
  \begin{pmatrix}
    a_{e, \mathrm{out}}\frac{\displaystyle e^{-ik_e x}}{\displaystyle \sqrt{k_e}}\\[0.4cm]
    a_{h, \mathrm{out}}\frac{\displaystyle e^{ik_hx}}{\displaystyle \sqrt{k_h}}
  \end{pmatrix}
  \label{eq:scatt}
\end{equation}
where $k_e=\sqrt{\mu+E}$ and $k_h=\sqrt{\mu-E}$ are the 
wave numbers of electrons and holes. Note that a hole propagating
to the right (incoming wave) is described by
$e^{-i k_h x}$.   In \eqref{eq:scatt} the outgoing coefficients $a_{e/h,\mathrm{out}}$
are related to the incoming coefficients by the scattering matrix $S(E)$
\begin{equation}
  \begin{pmatrix}
    a_{e,\mathrm{out}}\\[0.2cm]
    a_{h,\mathrm{out}}
  \end{pmatrix}
  =
  \begin{pmatrix}
    S_{ee}(E) & S_{eh}(E)\\[0.2cm]
    S_{he}(E) & S_{hh}(E)
  \end{pmatrix} 
  \begin{pmatrix}
    a_{e,\mathrm{in}}\\[0.2cm]
    a_{h,\mathrm{in}}
  \end{pmatrix}\ .
\end{equation}
The scattering matrix can be obtained by 
matching the solution in the normalconducting
part with a decaying solution in 
the superconducting part at $x=0$. 
The regime $E\ll \Delta_0 \ll \mu=k_F^2$ 
will be relevant to us and simplifies the
expressions considerably.
Let us formalise this regime by
using the
observation that the difference of the 
electron wave number
and the Fermi wave
number is proportional to the energy
$k_e-k_F=\sqrt{k_F^2+E}-k_F=
\frac{E}{2 k_F} +\mathcal{O}(E^2/k_F^3)$.
Keeping this difference fixed 
as $k_F \rightarrow \infty$ defines an 
interesting asymptotics. 
We thus introduce
$\kappa=\frac{E}{2 k_F}$ as the 
new (rescaled) energy
variable.
All other system parameters have to be 
rescaled appropriately.
For the pair potential we write
$\Delta_0= k_F^{1+\alpha} \delta^{1-\alpha}$
where $\epsilon<\alpha<1-\epsilon$ which ensures
$E\ll \Delta_0 \ll \mu$.
For the potential barrier $V_0=k_F v_0$ 
is the only way to scale the strength 
such that
the resulting limiting scattering problem 
still contains 
a finite barrier. 
The
leading order $S_A $ of the scattering matrix 
as $k_F\rightarrow \infty$
(keeping $\kappa$, $\delta$ and $v_0$ fixed)
only depends on the 
potential barrier strength $v_0$, explicitly 
\begin{equation}
  S_A=
  \begin{pmatrix}
    -\sqrt{1-t^2} e^{i \beta} & -i t \\
    -i t & -\sqrt{1-t^2} e^{-i \beta}
  \end{pmatrix}
\end{equation}
where
\begin{equation}
  t = \frac{2}{2+v_0^2} \qquad 
  e^{i\beta}= \frac{v_0+2i}{\sqrt{v_0^2+4}}\ .
\end{equation}
If there is no potential barrier $v_0=0$
the scattering matrix $S_A$ describes
a pure Andreev reflection where an incoming 
electron (hole) is
reflected as a hole (electron) and a phase 
factor $S_{he} \rightarrow -i $ is acquired.
In the opposite limit of a large potential 
barrier $v_0 \rightarrow \infty$
one has pure electron-electron 
(hole-hole) scattering 
with $S_{ee}, S_{hh} \rightarrow
-1$ which is equivalent to Dirichlet 
boundary conditions.\\
The leading asymptotics 
for the wave function
\eqref{eq:scatt} is obtained by replacing
replacing $S \rightarrow S_A$ and
$\frac{
  e^{\pm i k_{e,h}x}}{\sqrt{k_{e,h}}} 
\rightarrow
\frac{
  e^{\pm i (k_F\pm \kappa)x}}{\sqrt{k_{F}}} 
$. 
Note that this contains fast 
oscillations $\propto e^{i k_F x}$.
In the following we will 
always assume that $k_F$ is finite 
but large enough that it is justified
to consider the 
leading asymptotic term only. 

\subsection{Excitation spectrum of a 
  finite lead with one interface to a superconductor}

We can now discuss the discrete excitation 
spectrum 
$\{\kappa_n= \frac{E_n}{2k_F}\}$ of a finite
normalconducting lead of length $L$ ($-L<x<0$)
which is attached to a superconductor at $x=0$
and to an insulator at the other end $x=-L$. 
We assume Dirichlet boundary conditions
$\boldsymbol\psi_{\mathrm{NC}}(x=-L)=0$
at the interface to the insulator. 
Let us define the quantum map
\begin{equation}
  \mathcal{U}(\kappa)= S_A T(\kappa) S_D T(\kappa)
\end{equation}
where 
\begin{equation}
  S_D=
  \begin{pmatrix}
    -1 & 0 \\
    0 & -1
  \end{pmatrix} \qquad \text{and} \qquad
  T(\kappa)=
  \begin{pmatrix}
    e^{i(k_F+\kappa)L} & 0\\
    0 & e^{-i(k_F-\kappa)L}
  \end{pmatrix}\ .
\end{equation}
Here $T(\kappa)$ contains the phase factors
acquired by the electron and hole components
when propagated along
the lead from one side to the other. The quantum map contains four
steps: \textit{i.} 
propagation from the superconducting interface 
to
the insulating interface with $T(\kappa)$,  
\textit{ii.} scattering from the insulator 
with
$S_D$, \textit{iii.} propagation 
back to the
superconducting interface, and 
\textit{iv.} scattering from the
superconducting interface with $S_A$. 
If the quantum map 
$\mathcal{U}(\kappa)$ has an eigenvalue unity
then $\kappa$ belongs to the excitation 
spectrum.
The quantum map for the lead is periodic
$\mathcal{U}(\kappa+ \frac{\pi}{L} )
=\mathcal{U}(\kappa )$ -- the spectrum has 
the same periodicity and
it suffices to know the first two positive 
eigenvalues $\kappa_1$ and $\kappa_2$.
Explicitly, the quantum map is given by 
$\mathcal{U}(\kappa)= e^{i 2L \kappa} \mathcal{U}_0$
with
\begin{equation} \qquad
  \mathcal{U}_0=
  \begin{pmatrix}
    \sqrt{1-t^2} e^{i(2k_F L + \beta)} & i t\ e^{-i2k_F L }\\
    i t\ e^{i2k_F L }&\sqrt{1-t^2} e^{-i2k_F L + \beta)}
  \end{pmatrix}
\end{equation}
and has the eigenvalues
\begin{equation}
  u_1=e^{i(2L\kappa-\phi)} \qquad \text{and} \qquad
  u_2=e^{i(2L\kappa+\phi)}
\end{equation}
where
\begin{equation}
  \phi = \mathrm{arccos}\left(\cos(2 Lk_F+\beta) \sqrt{1-t^2}\right)
  \in [0,\pi] \ .
  \label{eq:phi}
\end{equation}
The first two positive eigenvalues in the spectrum of
the lead are
$
  \kappa_1=\frac{\phi}{2L}$ and $\kappa_2=\frac{2\pi-\phi}{2L}$
We arrive at the corresponding density of states
\begin{equation}
  d(\kappa)=\sum_{n=-\infty}^\infty \left[
    \delta\left(\kappa- \frac{2\pi n+\phi}{2L}\right) 
    +
    \delta\left(\kappa- \frac{2\pi n-\phi}{2L}\right) 
  \right]\ .
\end{equation}
We can now consider the mean density of states
$\langle d(\kappa)\rangle_{k_F}$
averaged over different values of the Fermi 
wavelength $k_F$
where 
$\langle \odot \rangle_{k_F}= \lim_{K\rightarrow \infty}
\frac{1}{K} \int_{k_{0}}^{k_0+K}\ \odot\ dk_F$.
The result will only depend on the
barrier strength $v_0$ or, equivalently,
on
the probability 
$p_A=t^2$
to be Andreev scattered at the 
superconducting interface.
From \eqref{eq:phi} one observes
that 
$\phi$ is always
confined to the interval
$  \mathrm{arcsin}\ t  \le \phi \le \pi - \mathrm{arcsin}\ t
$
and every value in that interval is 
assumed for some $k_F$.
As a consequence the mean density of states 
contains gaps
of size $2\Delta_g$ where 
\begin{equation}
  \Delta_g= \frac{1}{2L} \mathrm{arcsin} \sqrt{p_A}
\end{equation} 
at $\kappa= n \pi$ ($n=0,\pm 1,\pm 2,\dots$).
%The left graph in Figure \ref{fig1} shows the dependence of the
%size of the gap on $p_A$.
%While the periodicity of this gap is a 
%consequence of the
%simplicity of our model which only 
%contains a single length the 
%appearance of a gap or a strong 
%suppression of the density
%of states at $\kappa=0$ will be a 
%property of all models 
%discussed in this paper.\\
\begin{figure}
  \includegraphics[width=0.8\textwidth]{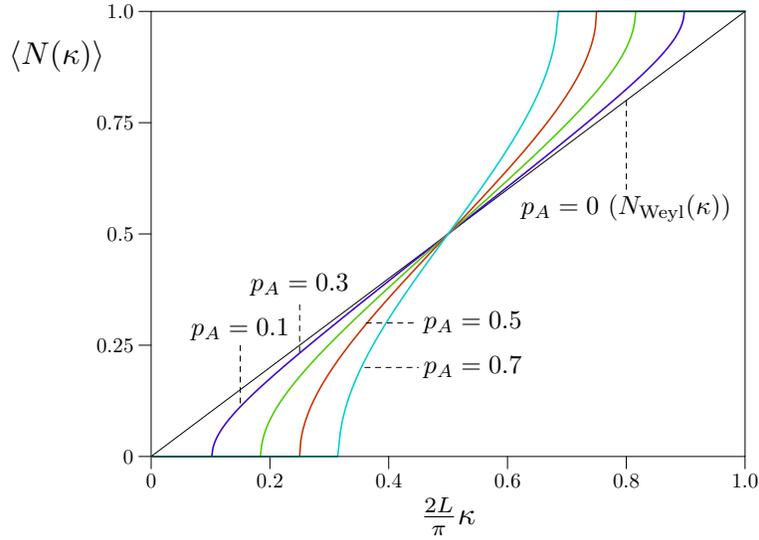}
  \caption{Averaged spectral counting function 
    for a single lead 
    (see text)
    for some values of the probability of 
    Andreev scattering $p_A$.
    The same plots also apply to homogeneous
    Andreev
    star graphs with $B$ bonds and negligible
    bond length fluctuations (see Section \ref{sec:homogeneous}).
  }
  \label{fig1}
\end{figure}
The average over the Fermi wavelength $k_F$ can 
be performed straight forwardly by
replacing $k_FL+\beta\rightarrow \theta$ and 
averaging over one period of $\theta$
\begin{equation}
  \begin{split}
    \left\langle d(\kappa) \right\rangle_{k_F}=&
    \begin{cases}
      0 & \text{for $0\le \kappa \le \frac{\mathrm{arcsin}\ {t}}{2L}$}\\
      \frac{2L}{\pi} \frac{\sin (2L \kappa)}{\sqrt{\sin^2 (2L\kappa)+t^2}} & 
      \text{for $\frac{\mathrm{arcsin}\ {t}}{2L}
        \le \kappa \le \frac{\pi-\mathrm{arcsin}\ {t}}{2L}$}\\
      0 & \text{for $\frac{\pi-\mathrm{arcsin}\ {t}}{2L}
        \le \kappa \le \frac{\pi }{2L}$}
    \end{cases}\\
    \left\langle d\left(\kappa+\frac{\pi}{2L}\right)
    \right\rangle_{k_F}=& 
    \left\langle d(\kappa) \right\rangle_{k_F}.
  \end{split}
  \label{eq:dos_B1}
\end{equation}
Figure \ref{fig1} shows the corresponding 
averaged spectral counting function
\begin{equation}
  \begin{split}
    \left\langle N(\kappa) \right\rangle_{k_F} 
    =& \int_0^\kappa d \kappa'
    \left\langle d(\kappa') \right\rangle_{k_F}\\
    =&
    \begin{cases}
      0 & \text{for $0\le \kappa \le \frac{\mathrm{arcsin}\ {t}}{2L}$}\\
      \frac{1}{\pi} \mathrm{arccos}\left(\frac{\cos (2L \kappa)}{\sqrt{1-t^2}}  \right) & 
      \text{for $\frac{\mathrm{arcsin}\ {t}}{2L}
        \le \kappa \le \frac{\pi-\mathrm{arcsin}\ {t}}{2L}$}\\
      1 & \text{for $\frac{\pi-\mathrm{arcsin}\ {t}}{2L}
        \le \kappa \le \frac{\pi }{2L}$}
    \end{cases}\\
    \left\langle N\left(\kappa+\frac{\pi}{2L}\right)
    \right\rangle_{k_F}=& 
    \left\langle N(\kappa) \right\rangle_{k_F}+1
  \end{split}
  \label{eq:cf_B1}
\end{equation}
for various values of the probability $p_A=t^2$
for Andreev scattering.

\section{The spectral gap in Andreev graphs}
\label{sec:gaps}

In this chapter we will generalise the 
discussion of the
lead with one superconducting/insulating
interface at the ends to Andreev star graphs.
A star graph $\mathcal{G}_B$ consists of one central vertex
and $B$ outer (or peripheral) vertices which 
are connected to 
the centre by $B$ bonds (or edges).
We assign a length 
$L_b$ ($b=1,\dots,B$) and a position 
variable $x_b\in [0,L_b]$ to each bond.
The star graph can then be quantised 
(and become a 
quantum star graph) by defining
a Schr\"odinger operator on the graph 
(see \cite{kuchment,graphreview} and references therein). 
An Andreev star graph can similarly be constructed
by defining a Bogoliubov-de Gennes operator on the graph.\\
Alternatively quantum graphs can be defined using the
scattering formalism of Kottos and Smilansky
\cite{kottos}.
Their approach has been used to construct 
Andreev graphs \cite{Andreevgraphs} 
(and other classes of quantum graphs that 
contain 
particle-hole symmetries). We will follow 
the latter 
approach and assume that the peripheral 
vertices of the 
star graphs have an interface to a 
superconductor 
such that Andreev reflections can occur with a certain 
probability. The rest of the 
graph (the bonds and the
central vertex) are normalconducting such that 
electron and hole components are not coupled.
The construction takes care that the symmetry 
$\boldsymbol{\psi}(x)=\begin{pmatrix} u(x)\\ v(x)
\end{pmatrix} \mapsto \mathcal{C} \boldsymbol{\psi}(x)
= \begin{pmatrix} - v(x)^* \\ u(x)^*
\end{pmatrix}$ which transforms positive energy states 
to negative energy states is also obeyed 
by the Andreev graph.
Indeed the lead that we have discussed in the previous 
section is already an example of such an Andreev 
star graph with $B=1$.
We will be interested what happens 
to the gap $\Delta_g$
for some ensembles of Andreev star graphs 
and how the
properties of the density of states depend on 
the properties of the central
vertex and the Andreev scattering amplitudes at
superconducting vertices.

\subsection{Construction of Andreev star graphs}
\label{sec:construction}

Andreev star graphs can be quantised using a 
quantum map $\mathcal{U}(\kappa)$ in a 
similar way
as we have treated the lead in the 
previous section.
For a star graph with $B$ bonds the 
quantum map is a $2B \times 2B$
matrix of the form
\begin{equation} 
  \mathcal{U}(\kappa)= S_A T(\kappa) 
  S_C T(\kappa)
\end{equation}
where 
\begin{equation}
  \begin{split}
    T(\kappa)=&
    \begin{pmatrix}
      e^{i(k_F+\kappa) \mathcal{L}} & 0\\
      0 & e^{-i(k_F-\kappa)\mathcal{L}}
    \end{pmatrix}\\
    \mathcal{L}=&
    \mathrm{diag}(L_1,\dots,L_B)
  \end{split}
\end{equation}
is a diagonal matrix which describes the propagation
of an electron (first $B$ entries) or hole
(last $B$ entries) from one end of each bond to the other ($\mathcal{L}$ is the diagonal $B \times B$ matrix
which contains the lengths of the bonds),
$S_A$
describes the scattering at the peripheral vertices
and $S_C$ the scattering from the central vertex.
The (partial) Andreev scattering at the peripheral vertices
has four blocks 
\begin{equation}
  S_A=
  \begin{pmatrix}
    - r & -i t\\
    -i t & -r
  \end{pmatrix}
\end{equation}
which contain the two diagonal matrices
\begin{equation}
  \begin{split}
    t=&\mathrm{diag}(t_1,\dots,t_B)\\
    r=&\mathrm{diag}(\sqrt{1-t_1^2},\dots, \sqrt{1-t_B^2})\ .
  \end{split}
\end{equation}
The real parameter $t_b\ge 0$ 
is the amplitude for Andreev
scattering at the $b$-th vertex. 
We will
not allow negative values for the
Andreev amplitude here\footnote{Andreev graphs 
with mixed signs for the Andreev scattering 
amplitude belong to a different universality 
class
\cite{Andreevgraphs} -- in an Andreev billiard this would 
correspond to having two interfaces to different 
superconductors whose pair potentials have a
relative phase $e^{-\pi}=-1$. No 
strong spectral gaps can be seen in 
such a regime.}.
Comparing this to the
scattering matrix at a superconducting interface in the
previous section one may observe that we omitted the phase factor 
$e^{\pm i \beta}$ in the electron-electron/hole-hole amplitude.
These turn out to be irrelevant once we average the
resulting spectra over the Fermi wave length (in the same way as
they have been irrelevant for the lead
-- here we omit them from the start).\\
The bonds and the central vertex are
assumed to be normalconducting. 
The electron-hole symmetry of the 
Bogoliubov-de Gennes equation and time-reversal
invariance
then restrict the scattering matrix of the 
central vertex to the form \cite{Andreevgraphs}
\begin{equation}
  S_C=
  \begin{pmatrix}
    \sigma & 0\\
    0 & \sigma^*
  \end{pmatrix}
\end{equation}
where $\sigma$ is a symmetric unitary 
$B \times B$ matrix.\\
We will write $\mathcal{G}_B\left(\mathcal{U}(\kappa)\right)$
for an Andreev star graph with $B$ bonds and quantum map $\mathcal{U}(\kappa)$.
The spectrum of the Andreev star graph consists of the discrete (positive and negative) values $\kappa=\kappa_n$
for which the quantum map has an eigenvalue unity. 
Equivalently, the spectrum is given by the zeros of the characteristic equation
\begin{equation}
  \mathrm{spec}\left[\mathcal{G}_B(\mathcal{U}(\kappa))\right]
  =
  \left\{ \kappa \in \mathbb{R} : \xi_{\mathcal{G}_B(\mathcal{U}(\kappa))}(\kappa,1)=0 \right\}
  \label{eq:spec}
\end{equation}
where
\begin{equation}
  \xi_{\mathcal{G}_B(\mathcal{U}(\kappa))}(\kappa,\lambda)=
  \mathrm{det}\left(1- \lambda\ \mathcal{U}(\kappa)\right)
  \label{eq:characteristic}
\end{equation}
The spectrum \eqref{eq:spec}
is symmetric around $\kappa=0$ -- if $\kappa_n$ is in the spectrum so is $-\kappa_n$.
This can be seen from the electron-hole symmetry
\begin{equation}
  \mathcal{U}(\kappa)^*=
  \begin{pmatrix}
    0 & 1\\
    -1 & 0
  \end{pmatrix}
  \mathcal{U}(-\kappa)
  \begin{pmatrix}
    0 & -1\\
    1 & 0
  \end{pmatrix}\ .
\end{equation}
It will be convenient to define the density of states as
\begin{equation}
  d(\kappa)=
  \frac{1}{B} \sum_n \delta(\kappa-\kappa_n)
\end{equation}
which differs from the conventional form by the factor $1/B$. Accordingly we define
the spectral counting function
\begin{equation}
  N(\kappa)=\int_0^\kappa d(\kappa') d\kappa'
  =\frac{1}{B}\sum_{\kappa_n >0 } \Theta(\kappa -\kappa_n)
\end{equation}
(where we have assumed that $\kappa=0$ is not in the spectrum and $\Theta(x)$ is the Heaviside step function).
The Kottos-Smilansky trace formula \cite{kottos}
expresses the spectral counting function (and
hence the density of states) in terms of the characteristic equation
\begin{equation}
  \begin{split}
    N(\kappa)=&N_{\mathrm{Weyl}}(\kappa)+ N_{\mathrm{osc}}(\kappa)\\
    N_{\mathrm{Weyl}}(\kappa)=&  \frac{2\mathrm{tr}(\mathcal{L})}{B \pi}\kappa\\
    N_{\mathrm{osc}}(\kappa)=&- \frac{1}{\pi B}\lim_{\lambda \rightarrow 1^-}
    \mathrm{Im}\,\mathrm{ln} \, \xi_{\mathcal{G}_B(\mathcal{U}(\kappa))}(\kappa, \lambda)\ .
  \end{split}
  \label{eq:trace_formula}
\end{equation}
The contribution $N_{\mathrm{osc}}(\kappa)$ is an oscillating function of 
the spectral parameter $\kappa$ while $N_{\mathrm{Weyl}}(\kappa)$
describes the mean increase of the number of states over a large
interval of the (rescaled) energy $\kappa$. 
The oscillating part of the counting function does not necessarily
vanish if averaged over some system parameter at fixed $\kappa$. 

\subsection{Andreev star graphs with negligible bond length fluctuations}

Let us introduce the bond length fluctuations by writing
\begin{equation}
  L_b=L +\delta L_b \qquad \text{where $\sum_b \delta L_b=0$}
\end{equation} 
where $L$ is the mean bond length. If the fluctuations $\delta L_b$ are very small, 
and for small enough $\kappa$ one has $e^{iL_b \kappa} \approx e^{i L \kappa}$.
Indeed we will only be interested in the spectrum
on a scale $|\kappa| \lesssim C \frac{\pi}{2 L}$ (such that $N_{\mathrm{Weyl}}(\kappa) 
\lesssim C$ and $C$ is of order unity) and assuming
$\delta L_b \frac{\pi}{2L} \ll 1/C$ 
we will set
$e^{i\delta L_b \kappa} \mapsto 1$.  
As a consequence 
$\mathcal{U}(\kappa+\frac{\pi}{L})=\mathcal{U}(\kappa)$ is periodic (and hence is the spectrum).
Also, the spectrum and eigenfunctions of the 
Andreev graph can be constructed 
straight forwardly from
the $2 B$ eigenvalues and eigenvectors of the quantum map $\mathcal{U}(\kappa)$
at $\kappa=0$.\\
The small fluctuations in the bond lengths can however not be neglected in the
phase factors containing the Fermi wavenumber $k_F$ -- indeed we want to
average spectral functions over arbitrary large values of the Fermi wavenumber.
A further significant simplification is obtained by assuming that all
bond length are rationally independent. 
By setting $e^{ik_F L_b} \mapsto e^{i \theta_b}$ we may then replace an average over
the Fermi wavelength by an average over a $B$-torus
\begin{equation}
  \begin{split}
    \langle \mathcal{F}(e^{i k_F L_1},\dots,e^{i k_F L_B}) \rangle_{k_F}=&
    \langle \mathcal{F}(e^{i \theta_1},\dots,e^{i \theta_B}) \rangle_{\theta}\\
    =& \frac{1}{(2\pi)^B} \int d^B \theta \mathcal{F}(e^{i \theta_1},\dots,e^{i \theta_B}) \ .
  \end{split}
\end{equation}
Writing $\boldsymbol\theta= \mathrm{diag} (\theta_1,\dots,\theta_B)$
we arrive at the following class of ensembles of quantum maps
for an Andreev star graph with $B$ bonds and negligible bond length fluctuations
\begin{equation}
  \begin{split}
    \mathcal{U}(\kappa)=&
    \begin{pmatrix}
      -r &-i t\\
      -i t & -r
    \end{pmatrix}
    \begin{pmatrix}
      e^{i \boldsymbol\theta} e^{iL \kappa} & 0\\
      0 & e^{-i \boldsymbol\theta} e^{iL\kappa}
    \end{pmatrix}
    \begin{pmatrix}
      \sigma & 0 \\
      0 & \sigma^*
    \end{pmatrix}
    \begin{pmatrix}
       e^{i \boldsymbol\theta} e^{iL \kappa} & 0\\
      0 & e^{-i \boldsymbol\theta} e^{iL\kappa}
    \end{pmatrix}\\
    =& - e^{i 2 L \kappa}
     \begin{pmatrix}
      r &i t\\
      i t & r
    \end{pmatrix}
    \begin{pmatrix}
      e^{i \boldsymbol\theta}\sigma e^{i \boldsymbol\theta} & 0 \\
      0 & e^{-i \boldsymbol\theta} \sigma^* e^{-i \boldsymbol\theta}
    \end{pmatrix}
  \end{split}
\end{equation} 
where $\theta_1,\dots,\theta_B$ are $B$ independent random variables
that are uniformly distributed over the interval $[0,2\pi)$. The remaining parameters
such as the central electron-electron scattering matrix $\sigma$ and the
Andreev scattering amplitudes $t_b$ are assumed to be fixed in one ensemble.
We will call the ensemble homogeneous if all Andreev scattering amplitudes are
equal $t_b=t$, otherwise we call it inhomogeneous.
We will now solve the homogeneous case and discuss some 
inhomogeneous ensembles numerically. 

\subsubsection{The homogeneous ensemble}
\label{sec:homogeneous}

In this case the ensemble averaged
spectral counting 
will turn out to be
completely independent of the choice 
of the central
electron-electron scattering matrix, and also of the number of bonds.
In fact the case can be reduced 
to the case $B=1$ that we have already solved.
To see this we 
write $\theta_b=\theta+\delta \theta_b$
such that $\sum \delta \theta_b =0$. We will first do the
average over $\theta$ and 
observe that this  can be solved by 
diagonalisation.
The straight forward result of the average 
over $\theta$ 
will not
depend on $\delta \theta_b$.\\
To do the 
average over $\theta$ let us fix $\boldsymbol{\delta}\boldsymbol{\theta}$ and
diagonalise
\begin{equation}
  e^{i \boldsymbol\delta \boldsymbol\theta} \sigma e^{i \boldsymbol\delta \boldsymbol\theta}=
  O e^{i \boldsymbol{\nu}} O^T
\end{equation} 
where $O$ 
is a real orthogonal matrix 
(remember that $\sigma$ is symmetric)
and $e^{i\boldsymbol{\nu}}=\mathrm{diag}(e^{i \nu_1},\dots,e^{i\nu_B})$ contains
the unimodular eigenvalues on the diagonal (we will omit the explicit dependence 
of the orthogonal matrix and the eigenvalues on $\delta \theta_b$).  
We may now write
\begin{equation}
  \mathcal{U}(\kappa)=
  \begin{pmatrix}
    O & 0\\
    0 & O
  \end{pmatrix}
  \begin{pmatrix}
    r & it\\
    i t & r
  \end{pmatrix}
  \begin{pmatrix}
    e^{i (2\theta + 2 \kappa L +\pi+\boldsymbol\nu) } & 0\\
    0 &  e^{-i (2\theta - 2 \kappa L +\pi+\boldsymbol\nu) }
  \end{pmatrix}
  \begin{pmatrix}
    O^T & 0\\
    0 & O^T
  \end{pmatrix}
\end{equation}
and observe that the spectral counting function can be written as a sum
over $B$ modes each of which is equivalent to the single lead (the case $B=1$).
As a consequence the position and size of the spectral gaps are the same as
for the single lead and the
 averaged density of states and the 
averaged spectral counting function
are described by \eqref{eq:dos_B1} and \eqref{eq:cf_B1} respectively (any 
dependence on the $\delta \theta_b$ has dropped).\\
It is remarkable that this result is universal -- it only depends
on the Andreev scattering amplitude and on no other system parameter
(the dependence on the length $L$ is trivial).

\subsubsection{Some inhomogeneous ensembles}

Different Andreev scattering amplitudes at the 
peripheral vertices introduce
an additional amount of disorder in the 
inhomogeneous case. One may expect
that the form of the averaged density of 
states now depends on the actual
choice of amplitudes $t_b$ 
\emph{and} on the central electron-electron
scattering matrix $\sigma$ -- and the 
same applies to the existence and
size of spectral gaps. In this section we 
will only discuss the extreme case
$t_b \in \{0,1\}$ -- that is, at some 
peripheral vertex one has either
complete Andreev reflections $t_b=1$ or complete normal reflections
$t_b=0$. Let $M\le B$ be the number of bonds with total
Andreev reflection such that $t_b=0$ for $b=1,\dots B-M$ 
and $t_b=1$ for $b=B-M+1,\dots,B$.
The mean probability to be Andreev scattered is
$p_A=\frac{M}{B}$.\\
One consequence of our choice of Andreev scattering 
amplitudes is that averaging over the phases $\theta_{B-M+1},\dots,
\theta_{B}$ is completely irrelevant. 
To see this let us assume an electron is scattered into 
one of the bonds $b=B-M+1,\dots,B$. It accumulates a phase 
$e^{i(\theta_b+\kappa L)}$ as it propagates 
from the centre to the
peripheral vertex where it is scattered back as a hole and acquires a further
phase factor $-i$. When the hole has propagated back to the centre
the overall phase which has been accumulated is $-i e^{2 \kappa L}$
where the contribution from the random phase $\theta_b$ has been cancelled 
exactly.\\
One further implication can be drawn 
from looking at the eigenvectors
of the matrix $\tilde\sigma(\boldsymbol\theta )=
e^{i\boldsymbol\theta}\sigma e^{i \boldsymbol\theta}$
where $\boldsymbol\theta=\mathrm{diag}(\theta_1,\dots,
\theta_B)$ is fixed. In general it is not possible to 
construct eigenstates of the Andreev star graph from the eigenvectors  $\tilde{\sigma}$ because
these get coupled in a different way at normal 
reflecting and at Andreev reflecting peripheral
vertices. However, if $\tilde{\sigma}$
has an eigenvector 
$\mathbf{a}=(0,\dots,0,a_{B-M+1},\dots,a_B)^T$
with eigenvalue $e^{i \alpha}$
which is supported completely
on the bonds $b=B-M+1,\dots,B$ then it is 
straight forward to show that $\mathcal{U}(0)$
has the two eigenstates $\begin{pmatrix}
  e^{-i\alpha/2} \mathbf{a}\\
  \pm e^{i\alpha/2} \mathbf{a} 
\end{pmatrix}$ which are also supported
on the bonds $b=B-M+1,\dots,B$. The 
corresponding eigenvalue is $\mp i$ which 
contributes to the spectrum of 
the Andreev graph at $\kappa \frac{2 L}{\pi}=
n+\frac{1}{2}$ ($n$ integer). 
Furthermore, if there is such a 
state for some fixed choice of 
$\boldsymbol\theta$ then there will be such a 
state for any element in the ensemble and one 
will observe a corresponding step in the 
averaged counting function.
A similar statement does not hold for states
that are supported on the normalconducting bonds $b=1,\dots,B-M$.

In the following we will restrict our discussion to two choices for
the central electron-electron scattering matrix
that are very well studied in the context of
normal quantum graphs:\\
\textit{i.} the Neumann 
(or Kirchhoff) scattering matrix
\begin{equation}
  \sigma_{\mathrm{Neumann}\, bb'}= \frac{2}{B} - \delta_{bb'}
  \label{eq:Neumann}
\end{equation}
and \textit{ii.}
the discrete Fourier transform (DFT) matrix
\begin{equation}
  \sigma_{\mathrm{DFT}\, bb'}=
  \frac{1}{\sqrt{B}} e^{i2\pi\frac{bb'}{B}}\ .
\end{equation}\begin{figure}
  \includegraphics[width=0.8\textwidth]{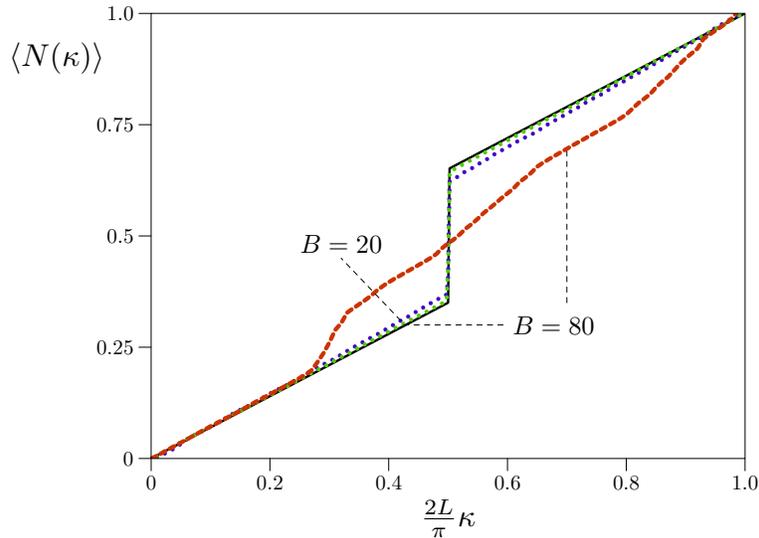}
  \caption{Dotted and full lines: 
    averaged spectral counting function for Andreev graphs with negligible 
    bond length fluctuations, inhomogeneous Andreev scattering, and a Neumann e-e scattering matrix. 
    The two dotted lines are for 
    $B=20$, $M=6$ (blue) and for $B=80$, $M=24$ (green) such that $p_A=0.3$ in both cases. 
    The full (black) line
    is the expected limiting function and hardly differs from the line for $B=80$.\newline
    Dashed (red) line: averaged spectral counting function for 
    the same inhomogeneous ensemble but with one set of
    randomly chosen bond lengths.
    }
  \label{fig2}
\end{figure}
For quantum star graphs (with a scalar wave function and with incommensurable bond lengths) 
it is known that the DFT matrix generates universal spectral statistics as known
from quantum chaotic systems and the Gaussian orthogonal ensemble (GOE) of
random matrix theory.
In contrast, the Neumann star graph belongs to a different universality class with
intermediate spectral statistics.\\
We first discuss the inhomogeneous ensemble with
a Neumann e-e
scattering matrix
$\sigma_{\mathrm{Neumann}}$.
The spectral decomposition of $\sigma_{\mathrm{Neumann}}$ is straight forward. It has one eigenvalue $1$ and $B-1$ eigenvalues $-1$. 
For our discussion it is important that one can 
construct $M-1$ orthogonal eigenvectors 
of $\sigma_{\mathrm{Neumann}}$ that are supported on the bonds $b=B-M+1,\dots,B$.
According to our discussion above this
leads to $2(M-1)$ eigenvectors 
of the quantum map $\mathcal{U}(0)$
with eigenvalues $\pm i$
and the averaged spectral counting function will 
have steps of size $\frac{M-1}{B}= p_A -\frac{1}{B}$ at $\kappa \frac{2 L}{\pi}
=\pm\frac{1}{2}+ n$ ($n=0,\pm 1, \pm 2, \dots$). On each of these modes
the probability to be Andreev scattered is unity. The remaining $2(B-M+1)$
modes are hardly coupled (at least for large $B$) by Andreev scattering -- the mean 
probability to be 
Andreev scattered among these modes is of the order $\frac{1}{B}$ and we may expect
that the effect on these modes is negligible in the limit 
$B\rightarrow \infty$ (where $p_A=\frac{M}{B}$ is fixed) such that the
contribution
to the averaged spectral counting function 
is linear. Figure
\ref{fig2}
compares the numerically obtained averaged counting function for
an Andreev graph with a Neumann e-e scattering matrix
with the expected limiting function 
\begin{equation}
  N(\kappa)_{\mathrm{Neumann}}= (1-p_A) \kappa \frac{2L}{\pi} + \sum_{n=-\infty}^\infty
  p_a \Theta \left(\kappa \frac{2L}{\pi} -\frac{1}{2} -n\right) .
\end{equation}
Our expectation seems to be confirmed by the numerics and suggests
that while there remains no gap in the averaged density of states the latter
is reduced compared to Weyl's law near $\kappa =0$ (modulo periodicity).\\
Let us now look more closely at the spectrum near
$\kappa=0$ for a finite size of the graph. Figure \ref{fig3}
shows  the integrated distribution $I_1(\kappa)$ of the first positive eigenvalue of the 
Andreev graph as a function of $\frac{2 L B}{\pi} \kappa$ (that is we measure the spectrum in units of the mean level spacing according to Weyl's law).
\begin{figure}
  \includegraphics[width=0.8\textwidth]{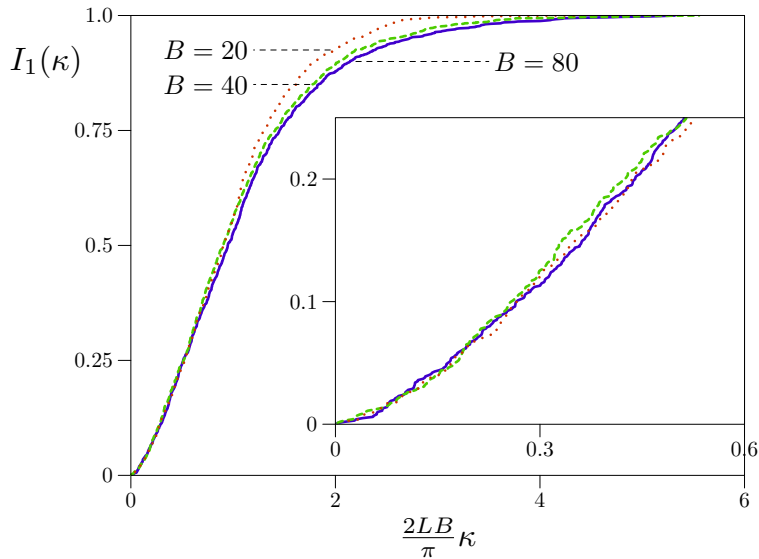}
  \caption{Integrated Distribution of the first positive eigenvalue of three
    Andreev star graphs with negligible bond length fluctuations, inhomogeneous 
    Andreev scattering ($p_A=0.3$ for all curves) and Neumann e-e
    scattering matrix. The three plots correspond to an increasing number of bonds 
    ($B=20,40,80$). }
  \label{fig3}
\end{figure}
\begin{figure}
  \includegraphics[width=0.8\textwidth]{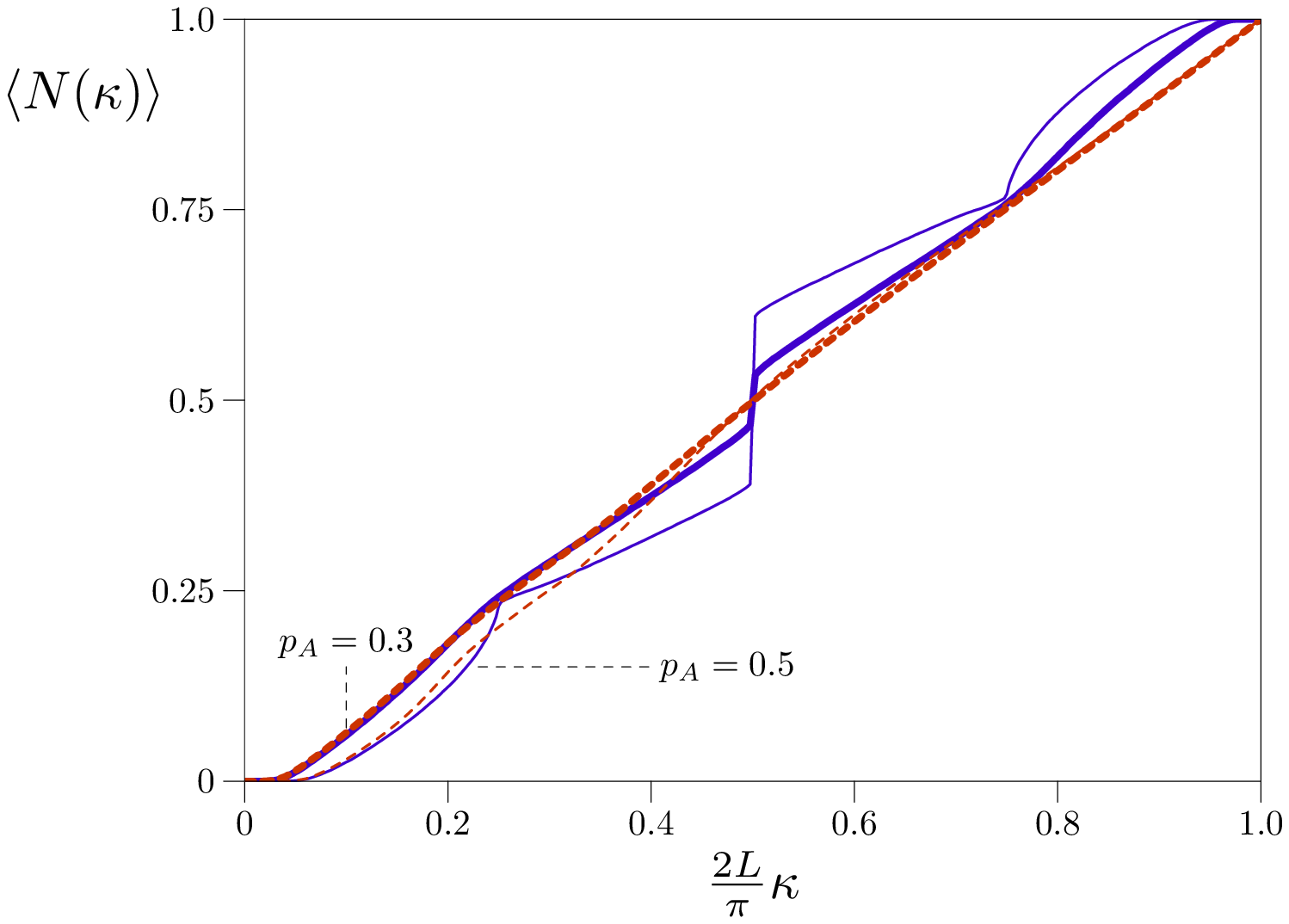}
  \caption{Full (blue) lines: 
    averaged spectral counting function for Andreev graphs with negligible 
    bond length fluctuations, inhomogeneous Andreev scattering, and a DFT e-e scattering matrix. 
    The fat line corresponds to $B=80$, $M=24$ ($p_A=0.3$) and the thin line to
    $B=80$, $M=40$ ($p_A=0.5$). Plots for $B=40$ and the same values for $p_A$ (not shown) 
    are almost identical.\newline
    Dashed (red) lines: averaged spectral counting function for the
    same inhomogeneous ensemble but with one set of
    randomly chosen bond lengths.
  }
  \label{fig4}
\end{figure}
The numerics suggests that even on this blown-up scale there is no gap.
$I_1(\kappa)$ may either start algebraically $I_1 \sim \kappa^\gamma$
or as a power series.
In both cases it implies that there is a further reduction (in addition to the one discussed 
above) of the density of states
on this blown-up scale.
We used 1000 realisations in our numerics which is not
sufficient to resolve the region
$\frac{2LB}{\pi} \kappa < 0.01$ so our results are not fully conclusive.

\begin{figure}
  \includegraphics[width=0.8\textwidth]{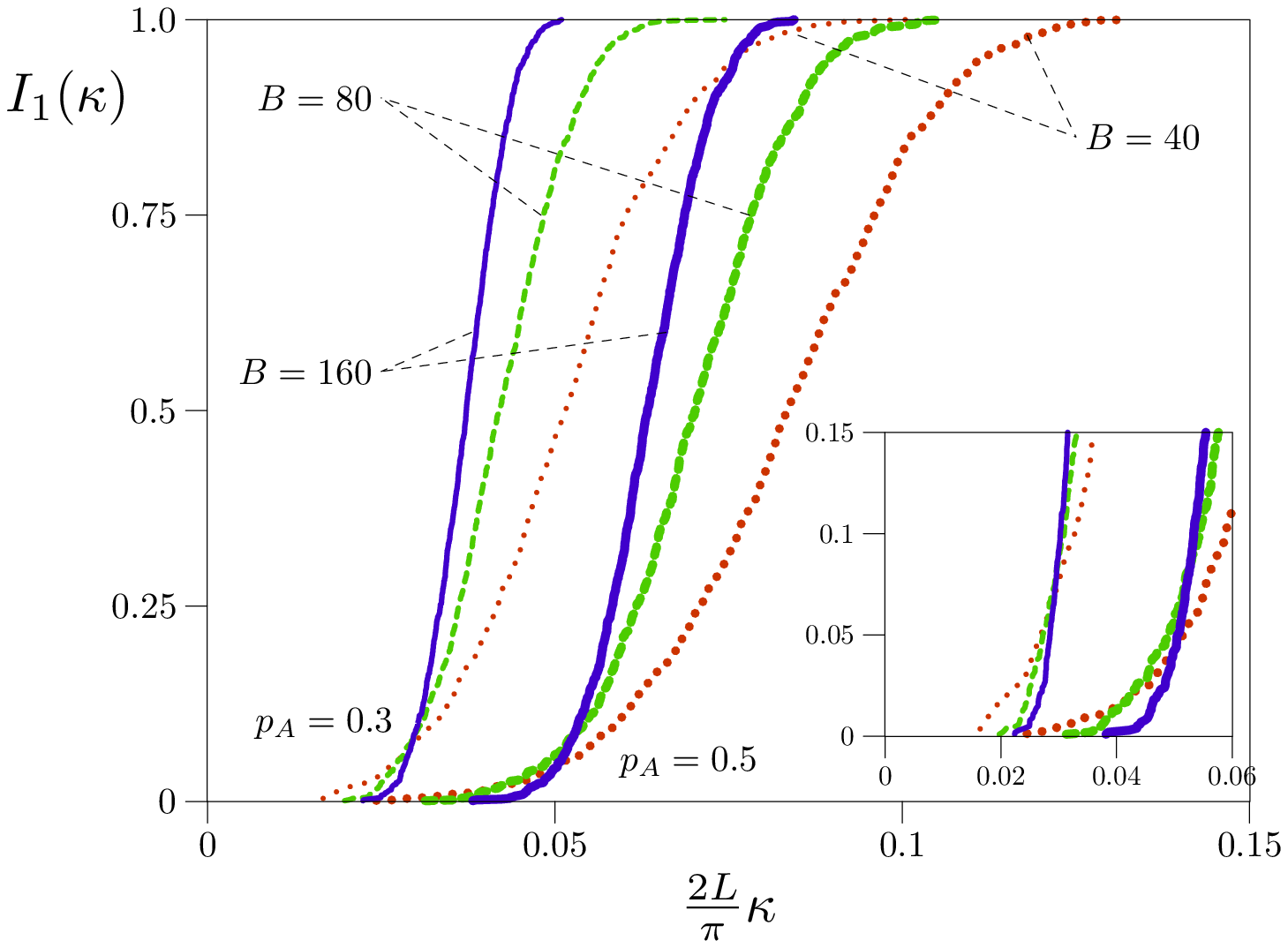}
  \caption{Integrated Distribution of the first positive eigenvalue of 
    Andreev star graphs with negligible bond length fluctuations, inhomogeneous 
    Andreev scattering and DFT e-e
    scattering matrix. The three plots on the left correspond to $p_A=0.3$, the three on the 
    right to $p_A=0.5$. In both cases we have plotted the distribution for $B=40$ (dotted red lines), $B=80$ (dashed green lines), and $B=160$ (full blue lines). }
  \label{fig5}
\end{figure}

The behaviour of the spectral counting function and $I_1(\kappa)$
is very different if we choose the DFT  e-e scattering matrix.
Figure \ref{fig4} shows the numerically averaged spectral counting function for this case.
There are several interesting features in this spectrum. Again, we find a jump at $\frac{2L}{\pi} \kappa=\frac{1}{2}$
which is connected to eigenvectors of $\sigma_\mathrm{DFT}$ which are supported on the bonds 
$b=B-M+1,\dots,B$ where the Andreev scattering occurs. However, the step is 
much smaller compared
to the Neumann case and the probability for Andreev scattering on 
the remaining modes is large enough
to have an effect. 
For large $B$ the spectral counting function converges (numerically) to a limiting
function -- the plots in Figure \ref{fig4}
are for $B=80$ and hardly differ from the case $B=40$ (not shown). Most importantly
the counting function seems to drop to zero at a finite positive value of $\kappa$.\\
The numerically obtained
distribution of the first
eigenvalue $I_1(\kappa)$ in Figure \ref{fig5}
reveals that a spectral gap appears at least
in the limit $B \rightarrow \infty$ (at constant
$p_A$). 

\subsection{Andreev star graphs with random bond lengths}

Let us now discuss the influence of bond lengths and discuss ensembles of Andreev graphs
for which the bond length fluctuations cannot be ignored. For a graph with $B$ bonds let
us choose a fixed set of bond lengths in the interval $0<L_b<1$ and now average over the
Fermi  wave number $k_F$. The main effect of the different bond lengths is that one 
cannot get the spectrum of the graph from just diagonalising the quantum map at $\kappa=0$
and the spectral properties are no longer periodic. In fact one should expect that 
any features in the counting function which differ from Weyl's law  can only survive
near $\kappa=0$. \\
We have calculated the averaged counting function numerically using the trace formula
\eqref{eq:trace_formula}.
We chose the bond lengths independently in the interval $0\le L_b \le 1$
with a random number generator. All plots we show are for a single realisation of the bond
lengths (we have assured that the figures are generic by comparing to other choices).
Indeed Figures \ref{fig2} and \ref{fig4} show  that the averaged counting function 
for both the Neumann and the DFT   e-e scattering matrix has lost all special features 
when $\frac{2L}{\pi} \kappa$ is on the order of one.
For smaller values $\frac{2L}{\pi} \kappa \ll 1$ the fluctuations in the bond lengths are not
effective and one gets back to the case discussed before. In short, the bond length fluctuations
destroy some special features of the previous ensembles with negligible
bond length fluctuations. The bond length fluctuations
do not have a strong influence on the 
reduction of the density of states near $\kappa=0$ or the
existence and size of a gap, at least when $p_A \ll 1$ \footnote{This should be taken 
with care since in our model the bond length 
fluctuation are still moderate.}.

\section{Conclusions}

We have shown that the spectra
of Andreev star graphs have a variety of
special spectral properties that are
related to the introduction of Andreev 
scattering into a quantum graph.
There is a spectral gap in homogeneous ensembles of
Andreev graphs. Both the
spectral counting function and the spectral gap
are universal and
only depend on the probability 
of Andreev scattering $p_A$. 
\\
Inhomogeneous ensembles contain additional 
disorder and the universal results from
the homogeneous case are not applicable.
Instead there is a rich variety of 
phenomena that depend on the properties of the 
central scattering matrix. 
For some special choices we could show
numerically that the spectral counting 
functions converge to a limiting function as
$B\rightarrow \infty$.\\
This first numerical investigation indicates
that the appearance and size 
of a 
spectral gap is strongly related to the
localisation properties of the eigenstates
of the central e-e scattering matrix.

\end{document}